# First experimental observation of plasmonic photonic jet based on dielectric cube

Igor V. Minin[1,2], Oleg V. Minin[1,2], Dmitry S. Ponomarev[3,4,5], Igor A. Glinskiy[3,4,6],

Dmitry I. Yakubovsky[5], Valentin. S. Volkov[5]

[1]National Research Tomsk Polytechnic University, Tomsk 634050, Russia

[2]National Research Tomsk State University, Tomsk 634050, Russia

[3]Institute of Ultra High Frequency Semiconductor Electronics RAS, Moscow 117105, Russia

[4]Prokhorov General Physics Institute of the Russian Academy of Sciences, Moscow 119991, Russia

[5]Moscow Institute of Physics and Technology, Dolgoprudny 141700, Russia

[6]MIREA–Russian Technological University, Moscow 119454, Russia

ABSTRACT

We report on the first experimental demonstration that the microstructure based on a dielectric cuboid combined to a thin metal film can act as an efficient plasmonic lens allowing focusing surface plasmons in a subwavelength scale. By means of numerical simulation of surface plasmon-polariton (SPP) field intensity distributions in the microstructure, we observe a low divergence of plasmonic photonic jet (PPJ) and high intensity subwavelength spots at the communication wavelength of 1530 nm. Then we fabricate an experimental sample of the microstructure and study the SSP field intensity distributions using the amplitude and phase-resolved scanning near-field optical microscopy and demonstrate the experimental observation of the PPJ effect for the SSP waves. Such novel and simple platform can provide new pathways for plasmonics, high-resolution imaging, biophotonics as well as optical data storage.

It is well known, that the surface plasmon polaritons (SPP) waves are highly localized at metal–dielectric interface, and thus have shorter wavelength than excitation radiation [1,2,3]. Excitation and effective control of the SPP require different components but the crucial among them is a lens.

Different structures based on slits, holes, rings [4,5,6,7,8], mirrors [9] and metasurfaces [10] have been proposed to realize subwavelength focusing, but the obtained focal spot size was larger than a half of the wavelength [6,9]. The Maxwell's fish-eye and Luneburg SPP lenses [11,12] demonstrate the capability to control the focusing properties of a lens by selecting the height of the dielectric lens in order to obtain the optimal effective refractive index contrast. Dielectric structures [13,14] and plasmonic gradient-index metasurface Luneburg lens based on subwavelength holes in the dielectric thin film [15] and mounted on the metal film have become an efficient method for the focusing and manipulating of two-dimensional SPPs waves. But dielectric plasmonic lenses cannot achieve the manipulation of an optical energy below its wavelength due to diffraction [13,14,15].

On the other hand, in [16] it was theoretically shown that micro-scaled dielectric disks with refractive index of $n = 1.5$ and radius of $r = 3$ μm are able to produce the so-called plasmonic photonic jet (PPJ, analog of the photonic nanojet effect [17,18], but for plasmonic waves), when the disks are excited by the SPP at optical frequencies featuring a wavelength of 800 nm. Later, the formation of the PPJ was theoretically predicted in a dielectric micro cuboid particle [19]. The structure comprised a 100 nm thick gold film deposited on top of a dielectric substrate with refractive index $n = 1.5$. The three-dimensional (3D) dielectric cuboid of silicon nitride ($Si_3N_4$) with refractive index $n = 1.97$ and lateral dimensions $l_x = l_z = \lambda_0 = 1550$ nm (where $\lambda_0$ is a telecommunication wavelength) was placed on top of the metal film. It was demonstrated that the location of the PPJ can be tailored by changing the height of the dielectric particle [16,19]. It was also shown that the best performance at $\lambda_0$ is observed when the height of the dielectric cuboid is about $0.1\lambda_0$ that allows for producing of the PPJ with a resolution of $0.68\lambda_0$ and an intensity enhancement of ~5 at the focus [19]. At the same time, this mechanism of SPP field localization was used to combine with 3D $Si_3N_4$ cuboids featuring a chain of dielectric particles [20] in order to extend the SPP propagation up to several times [21,22].

However, to the best of our knowledge, no experimental study of the PPJ effect for SPP waves has been performed yet. In this paper we demonstrate realization and first experimental observation of the PPJ effect in the dielectric cuboid mounted on a metal film and provide its theoretical and experimental characterization.

We used a full-wave 3D simulation via a finite element method (COMSOL Multiphysics) to solve the SPP wave vector ($\beta$) and to demonstrate the PPJ effect that is produced by dielectric cuboid combined to gold metal film. The effective refractive (mode) index of the SPP wave $n_{eff}$ is defined as $n_{eff} = \beta/k_0$ [10,20], where $k_0$ is a wave number. In the simulation, the SPP wave is incident onto the flat cuboid surface.

The SPP was illuminated by an optical excitation with $\lambda_0 = 1530$ nm featuring maximum electric field of $E=1$ Vm$^{-1}$ at the dielectric–metal interface. To simulate the surface plasmons diffraction, we used perfectly matched layers conditions and a non-uniform mesh featuring a minimum cell size of $1/5\lambda_0$ at the dielectric-metal interface. All the field distributions shown below are normalized to its maximum values. Some other details of the simulation are given in [23].

For simplicity and accordance to experiment we used Poly(methyl methacrylate) (PMMA) featuring a refractive index of $n = 1.5302$ (at $\lambda_0 = 1530$ nm) as a material for the dielectric cuboid. It should be noted that despite the refractive index of $Si_3N_4$ is about $n = 1.97$ [17] (in contrast to the PMMA refractive index), the formation of PPJ via the dielectric cuboid has similar trends for these materials. The value of the refractive index has been selected to circumvent a dispersion model and to simplify the simulations (this can be done because the variation of $n$ is of the order of $10^{-2}$ within the spectral band of interest [24]). We used a Drude–Lorentz dispersion model in simulation with relative permittivity of metal (gold) equal to $\varepsilon_m = -114.47 + 8.51i$ at $\lambda_0 = 1530$ nm [25]. The effective refractive index of the SPP is determined by both dielectric and metal that form the interface and is derived by the relation [26]:

$$k_{spp} = k_0((\varepsilon_m \cdot \varepsilon_d)/(\varepsilon_m + \varepsilon_d))1/2 = n_{eff} \cdot k_0, \quad (1)$$

where $k_0 = 2\pi/\lambda_0$ is the wave number in vacuum, while $\varepsilon_m$ and $\varepsilon_d$ are the relative permittivities of metal and dielectric, respectively. It means that for our conditions, the surface plasmon wavelength is $\lambda_{spp} = 0.978\lambda_0 = 1497$ nm (or about 2% less than the wavelength in a free space).

The numerical results of the SPP field intensity distributions within the $xz$-plane for different dielectric cuboid heights ($h$ = 200, 250 and 300 nm) with the same lateral dimensions $l_x = l_z = 3.26\lambda_0 = 5$ μm at $\lambda_0 = 1530$ nm are shown in Fig. 1 (note that a 3D geometry of the microstructure is shown in Figure 5 below).

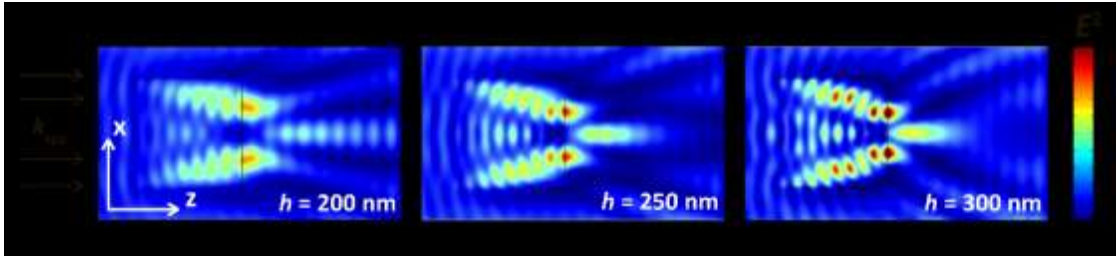

Figure 1. SPP field intensity $E^2$ distributions in the dielectric cuboid combined to metal film with different cuboid height of $h$ = 200, 250 and 300 nm. The cuboid is illuminated with $\lambda_0$ = 1530 nm. The surface plasmon wave propagates from left to right.

Figure 2 describes SPP field intensity $E^2$ distributions near the shadow surface of the dielectric cuboid in a focal spot (a) and along PPJ propagation (b). It is follows from Fig. 2b, that the PPJ demonstrates an exponential decay length in $z$ direction as $\exp(-k_d z)$, where the parameter $k_d$ is given by the dispersion relations [14,17]. From the comparison of Fig.2a and Fig.2b, one can mention that when the height of the cuboid alters from $h = 250$ nm to 350 nm, the degree of localization increases linearly: the full width at half maximum (FWHM) increases from $0.58\lambda_0$ to $0.77\lambda_0$ with a corresponding decrease in the PPJ length from $1.83\lambda_0$ to $1.51\lambda_0$. From Fig. 2a one can notice that the increase of the dielectric height from 200 nm to 350 nm leads to the intensity enhancement near the shadow surface of the cuboid and thus the effective refractive index $n_{eff}$ of the considered structure has a perceptible dependence on the height of the cuboid, as expected [12,14,17,19,23]. Moreover, according to Fig.2b, the maximal field intensity along direction of the SPP propagation has a red shift with the increase of the dielectric height.

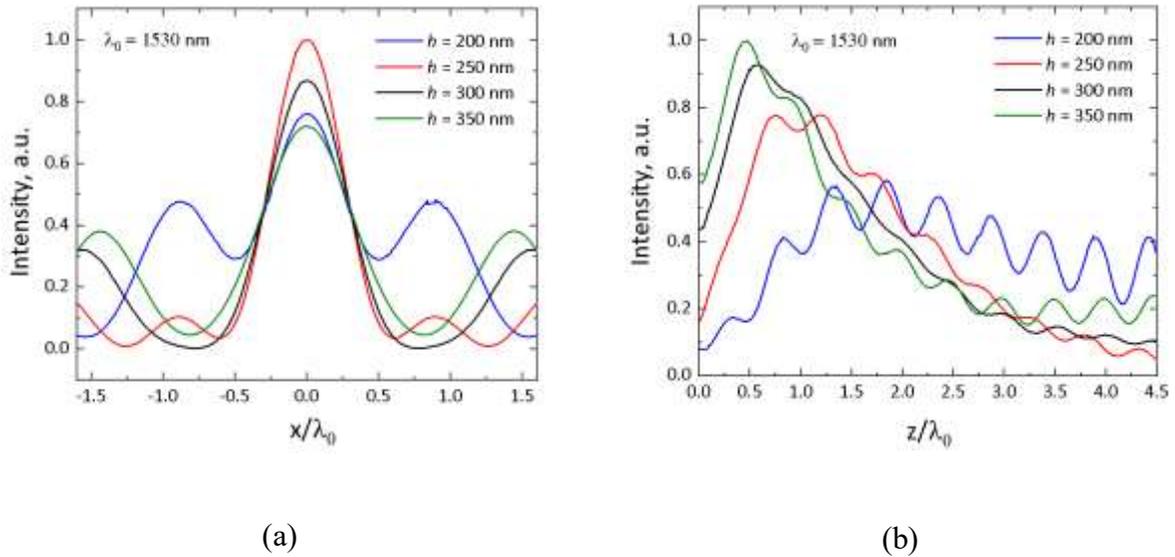

(a) (b)

Figure 2. SPP field intensity $E^2$ distributions near the shadow surface of the dielectric cuboid in a focal spot (a) and along PPJ propagation (b).

It is worthwhile noticing that the both parameters $h$ and $n_{eff}$ increase simultaneously thanks to the exponential decay of the SPP electric field within the cuboid, as expected [14]. With a further increase in the thickness of the cuboid, the monotonicity of the field intensity distribution along the propagation axis in the region of its localization is violated (Fig.2b). Therefore, according to the aforementioned results, we fixed $l_x = l_z = 3.26\lambda_0$ and $h = 250$ nm at $\lambda_0 = 1530$ nm.

The FWHM and the length of PPJ defined as a full length at half maximum (FLHM) of the PPJ propagation at different excitation wavelengths are shown in Fig.3. As seen, the transversal resolution FWHM of $0.58\lambda_0$ that lies above the diffraction limit of $0.5\lambda_0$ and the FLHM of the PPJ of 2.17 $\mu$m (i.e. $1.41\lambda_0$) are obtained at $\lambda_0 = 1530$ nm. Nevertheless, it is important to note that the FWHM value can be further decreased at optimized microstructure parameters with thus allowing focusing of SPP waves even below the diffraction limit [23].

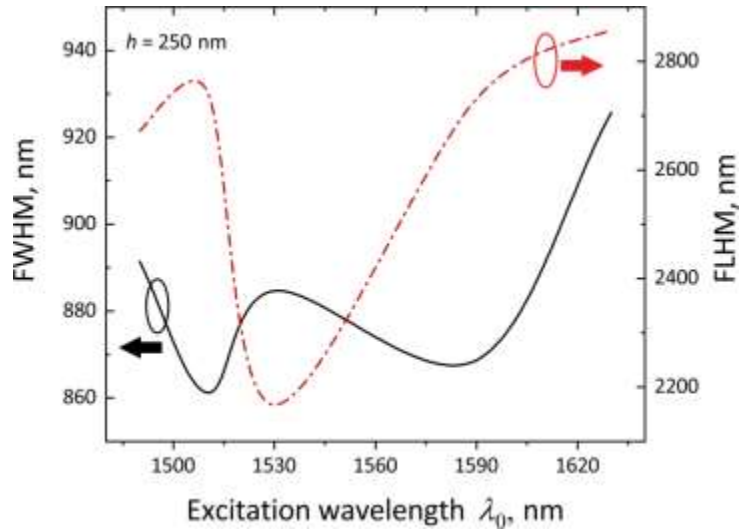

Figure 3. FWHM (solid curve) and FLHM (dash-dotted curve) of the PPJ propagation vs. the excitation wavelength.

The experimental samples of the dielectric cuboids were fabricated using a PMMA and then deposited on the surface of a 100-nm gold metal film. The height and the width of the cuboids were equal to $h = 250$ nm and $w = 5$ μm respectively. In order to excite the SPP under an optical pump of 1530 nm we fabricated a plasmonic grating comprising four 770 nm-wide grooves formed in the gold film with a period of 1540 nm featuring a filling fraction of 50%.

To demonstrate the realization of the PPJ effect for SPP waves we used a scattering-type scanning near-field optical microscopy (s-SNOM) [27]. The s-SNOM (NeaSNOM from Neaspec GmbH) is working as an atomic force microscope in a taping mode with a sharp metal-coated silicon tip, as near-field probe, oscillating at the resonance frequency of $\Omega \approx 280$ kHz with amplitude ~55 nm. In order to direct the SPP wave into the dielectric cuboid, plasmonic grating coupler was illuminated from below by a linearly polarized light at a normal angle to the sample surface (transmission configuration). Tunable telecom laser operating in the range of 1460–1640 nm (from Agilent) was used. In this s-SNOM arrangement while mapping of near-field and topography across the scan area of 30×10 μm$^2$, an illumination system remains aligned on plasmonic grating due to its synchronization with the sample moving. The tip-scattered light is collected by a top parabolic mirror and then is incident on the detector.

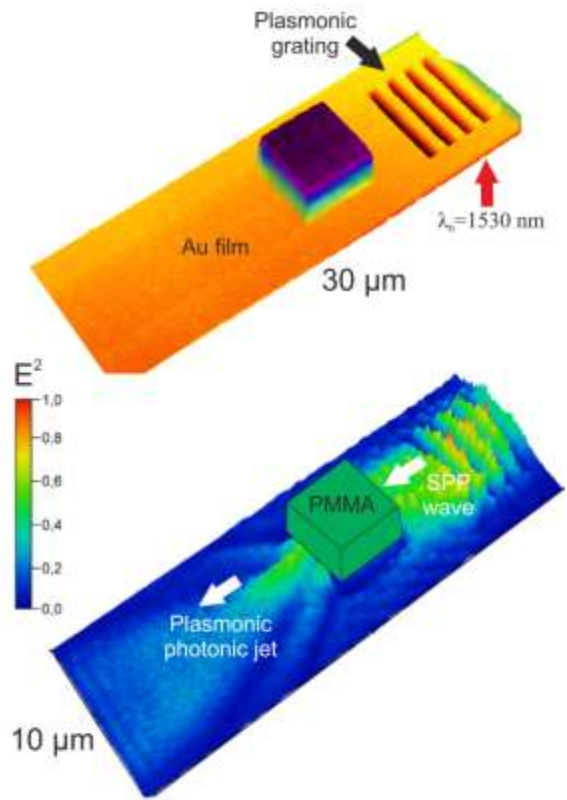

Figure 4. Topography image of the microstructure (dielectric cuboid on metal film) and PPJ propagation for the SPP waves derived using s-SNOM. The optical excitation of $\lambda_0 = 1530$ nm is incident on the backside of the plasmonic grating resulting in the excitation of SPP waves in the orthogonal direction.

To provide a clear imaging of the near-field distribution, most part of optical background was suppressed by demodulation of the detected signal at high-order harmonic frequency n$\Omega$ ($n$ = 2, 3, 4) and also by using interferometric pseudoheterodyne detection scheme with a modulated reference beam. In our case the signal of demodulation at the third harmonic (3$\Omega$) was taken into consideration, which was enough for background-free near-field analysis. The topography image of the microstructure (a) as well as the optical amplitude illustrating the experimental observation of PPJ propagation for the SPP waves under the optical excitation of $\lambda_0 = 1530$ nm (b) derived by s-SNOM are depicted in Fig.4. Note that for simplicity and demonstration, we used merely one excitation wavelength.

Figure 5 illustrates the experimental SPP field intensity $E^2$ distributions near the shadow side of the microstructure in a focal spot ($x$-direction) (a) and along PPJ propagation ($z$-direction) (b) demonstrating a reasonable agreement between modeling and experiment. As seen, the experimental FWHM value at $\lambda_0 = 1530$ nm is equal to $0.6\lambda_0$ (a) corresponding to the FLHM = 5.43 µm (i.e. $3.54\lambda_0$) (b). Comparing the obtained results with those for the modeling (see Fig.4), one can conclude that the localization of the experimentally observed SPP photonic nanojet along the propagation axis demonstrates a 2.5-fold enhancement compared to the modeling. Nevertheless, the interpretation of this phenomenon lies beyond the scope of this paper and will be the topic of further discussions.

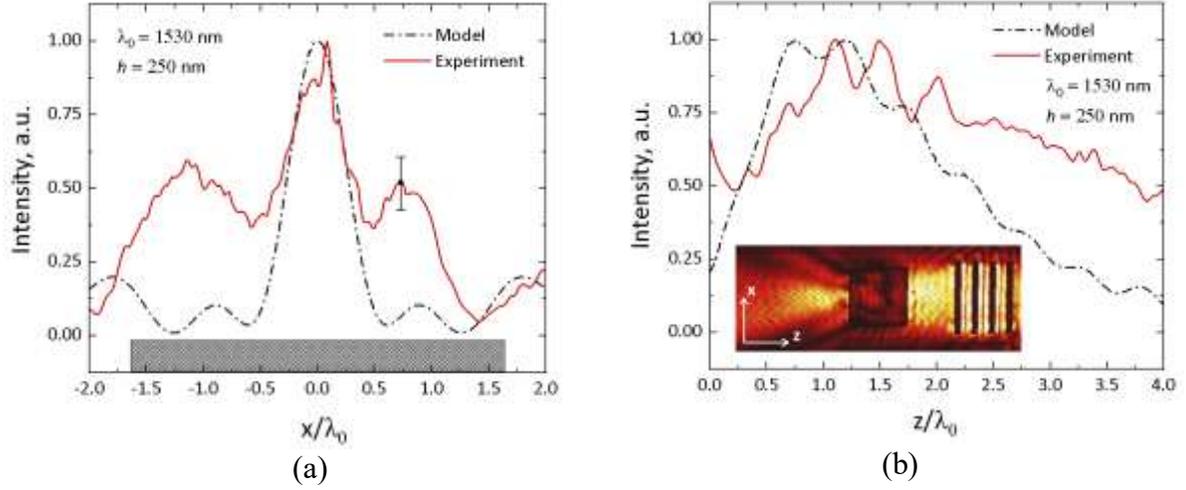

(a)                               (b)

Figure 5. SPP field intensity $E^2$ distributions near the shadow side of the dielectric cuboid mounted on metal film in a focal spot (*x*-direction) (a) and along PPJ propagation (*z*-direction) (b). The solid curves correspond to the modeling while the dash-dotted curves refer to the experiment. The cuboid borders are schematically shown by grey rectangle in (a), while the insert in (b) depicts the SNOM image of $E^2$ distributions. The excitation wavelength is $\lambda_0 = 1530$ nm.

In summary, we have experimentally demonstrated that a simple rectangular dielectric structure combined to a metal film can act as plasmonic lens [28] and thus can focus surface plasmons in a subwavelength scale. We observed a low divergence of plasmonic photonic jet (PPJ) and high intensity subwavelength spots at the communication wavelength of $\lambda_0 = 1530$ nm.

We performed numerical simulation using finite element method of surface plasmon-polariton (SPP) field intensity distributions in the cuboid featuring different geometry and deposited on gold metal film and observed the light which is scattered from the cuboid by imaging of the PPJ. Then we fabricated the experimental sample and demonstrated the first experimental observation of the PPJ effect for SPP waves by using a rather simple dielectric microstructure.

It is important to note that the PPJ with small subwavelength scale dimension is believed to show exciting potential applications in integrated and near-field optics to guiding and control of the subwavelength focusing of SPPs and a relatively cheap and simple way for the manipulation of the SPP propagation even below the diffraction limit at subwavelength scale (at optimized microstructure parameters).

Such novel and simple platform can provide new pathways for plasmonics, high-resolution imaging, biophotonics and optical data storage.

REFERENCES

1. Liu, Z.W.; Steele, J.M.; Srituravanich, W.; Pikus, Y.; Sun, C.; Zhang, X. Focusing Surface Plasmons with a Plasmonic Lens. Nano Lett. 2005, 5 (9), 1726–1729.

2. Lin, L.; Goh, X.M.; McGuinness, L.P.; Roberts, A. Plasmonic Lenses Formed by Two-Dimensional Nanometric Cross-Shaped Aperture Arrays for Fresnel-Region Focusing. Nano Lett. 2010, 10 (5), 1936–1940.

3. Kim, H.C.; Cheng, X. Gap Surface Plasmon Polaritons Enhanced by a Plasmonic Lens. Opt. Lett. 2011, 36 (16), 3082–3084.

4. Lindquist, N.C.; Nagpal, P.; Lesuffleur, A.; Norris, D.J.; Oh S.-H. Three-Dimensional Plasmonic Nanofocusing. Nano Lett. 2010, 10, 1369–1373.

5. Wu, G.; Chen, J.J.; Zhang, R.; Xiao, J.H.; Gong, Q.H. Highly Efficient Nanofocusing in a Single Step-like Microslit. Opt. Lett. 2013, 38 (19), 3776–3779.

6. Takeda, M.; Okuda, S.; Inoue, T.; Aizawa, K. Focusing Characteristics of a Spiral Plasmonic Lens. Jpn. J. Appl. Phys. 2013, 52, 09LG03.

7. Mote, R.G.; Minin, O.V.; Minin, I.V. Focusing behavior of 2-dimensional plasmonic conical zone plate. Opt. Quant. Electron. 2017, 49, 271.

8. Long, Y.; Zhang, Z.; Su, X. Robust subwavelength focusing of surface plasmons on graphene. EPL. 2016, 116, 37004.

9. Melentiev, P.N.; Kuzin, A.A.; Negrov, D.V.; Balykin, V.I. Diffraction-Limited Focusing of Plasmonic Wave by a Parabolic Mirror. Plasmonics. 2018, 13 (6), 2361–2367.

10. Yin, X.; Steinle T.; Huang, L.; Taubner, T.; Wuttig, M.; Zentgraf, T.; Giessen, H. Beam switching and bifocal zoom lensing using active plasmonic metasurfaces. Light: Science & Applications. 2017, 6, e17016.

11. Smolyaninova, V.N.; Smolyaninov, I.I.; Kildishev, A.V.; Shalaev, V.M. Maxwell fish-eye and Eaton lenses emulated by microdroplets, Opt. Lett. 2010, 35, 3396–3398.

12. Zentgraf, T.; Liu, Y.; Mikkelsen, M. H.; Valentine, J.; Zhang, X., Nat. Nanotechnol. 2011, 6, 151–155.

13. Hohenau, A.; Krenn, J. R.; Stepanov, A. L.; Drezet, A.; Ditlbacher, H.; Steinberger, B.; Leitner, A.; Aussenegg, F. R. Dielectric optical elements for surface plasmons, Opt. Lett. 2005, 30 (8), 893–895.

14. Shi, W.-B.; Chen, T.-Y.; Jing, H.; Peng, R.-W.; Wang, M. Dielectric lens guides in-plane propagation of surface plasmon polaritons. Opt. Express. 2017, 25 (5), 5772-5780.

15. Gartcia-Ortiz, C.E.; Cortes, R.; Gomez-Correa, J.E.; Pisano, E.; Fiutowski, J.; Garcia-Ortiz, D.A.; Ruitz-Cortes, V.; Rubahn, H.-G.; Coello V. Plasmonic metasurface Luneburg lens. Photonics Research. 2019, 7(10), 1112.

16. Ju, D.; Pei H.; Jiang, Y.; Sun, X. Controllable and enhanced nanojet effects excited by surface plasmon polariton. Appl. Phys. Lett. 2013, 102, 171109.

17. Heifetz, A.; Kong, S.C.; Sahakian, A.V.; Taflove, A.; Backman, V., Photonic Nanojets, Journal of Computational and Theoretical Nanoscience. 2009, 6/9, 1979-1992.

18. Luk'yanchuk, B.S.; Paniagua-Domínguez, R.; Minin, I.V.; Minin, O.V.; Wang Z. Refractive index less than two: photonic nanojets yesterday, today and tomorrow. Optical Materials Express. 2017, 7(6), 1820-1847.

19. Pacheco-Pena, V.; Minin, I.V.; Minin, O.V.; Beruete, M. Comprehensive analysis of photonic nanojets in 3D dielectric cuboids excited by surface plasmons. Ann. Phys. 2016, 528, 1–9.

20. Minin, I.V.; Minin, O.V.; Pacheco-Peña, V.; Beruete M. All-dielectric periodic terajet waveguide using an array of coupled cuboids. Appl. Phys. Lett. 2015, 106, 254102.

21. Pacheco-Pena, V.; Minin, I.V.; Minin, O.V.; Beruete, M. Increasing Surface Plasmons Propagation via Photonic Nanojets with Periodically Spaced 3D Dielectric Cuboids. Photonics. 2016, 3, 1–7.

22. Zhao, C.; Liu, Y.; Zhao, Y.; Fang, N.; Huang, T.J. A reconfigurable plasmofluidic lens. Nat. Commun. 2013, 4, 2305.

23. Minin, I.V.; Minin, O.V.; Ponomarev, D.S.; Glinskiy, I.A. Photonic Hook Plasmons: A New Curved Surface Wave. Ann. Phys. 2018, 530(12), 1800359.

24. Beadie, G.; Brindza, M.; Flynn, R.A.; Rosenberg, A.; Shirk, J.S. Refractive index measurements of poly(methyl methacrylate) (PMMA) from 0.4-1.6 μm, Appl. Opt. 2015, 54, F139-F143.

25. Yakubovsky, D.I.; Arsenin, A.V.; Stebunov, Y.V.; Fedyanin, D.Yu.; Volkov, V.S. Optical constants and structural properties of thin gold films. Optics Express. 2017, 25 (21), 25574.

26. Shi, W.-B.; Chen, T.-Y.; Jing, H.; Peng, R.-W.; Wang, M. Dielectric lens guides in-plane propagation of surface plasmon polaritons, Opt. Express. 2017, 25, 5772.

27. Fedyanin, D.Yu.; Yakubovsky, D.I.; Kirtaev, R.V.; Volkov, V.S. Ultralow-Loss CMOS Copper Plasmonic Waveguides, Nano Lett. 2016, 16, 1, 362-366.

28. Fu, Y.; Wang, J.; Zhang, D. (October 24th 2012). Plasmonic Lenses, Plasmonics - Principles and Applications, Ki Young Kim, IntechOpen, DOI: 10.5772/50029. Available from: https://www.intechopen.com/books/plasmonics-principles-and-applications/plasmonic-lenses

AUTHOR INFORMATION

**Corresponding Authors**

*Igor V. Minin, E-mail: prof.minin@gmail.com

* Dmitry S. Ponomarev, E-mail: ponomarev_dmitr@mail.ru

**Author Contributions**

I.V.M and O.V.M conceived and supervised the project, wrote the paper. I.A.G performed the numerical simulations. D.S.P wrote and edited the paper. R.M, D.I.Y and V.S.V fabricated the experimental samples and provided SNOM measurements. All authors contributed to discussions and editing the article.

**Funding**

I.A.G and D.S.P acknowledge the financial support of the Russian Scientific Foundation (Project No.18-79-10195), I.V.M. and O.V.M were partially supported by the Russian Foundation for Basic Research (Grant No. 20-57-S52001). The work was partially carried out within the framework of the Tomsk Polytechnic University Competitiveness Enhancement Program, Russia.